# ELECTRON-CLOUD EFFECTS IN PAST & FUTURE MACHINES – WALK THROUGH 50 YEARS OF ELECTRON-CLOUD STUDIES

F. Zimmermann, CERN, Geneva, Switzerland

*Abstract*

Past electron-cloud (e-cloud) observations, studies and mitigation techniques are quickly reviewed along with some ongoing code developments, the preceding ECLOUD workshops, recent contacts with the spacecraft community, the important role of Francesco Ruggiero, and a few current electron-cloud topics discussed at ECLOUD12 in La Biodola.

## HISTORICAL ENCOUNTERS

Starting in 1965 on a small proton storage ring at BINP, the Argonne ZGS, and the BNL AGS, and continuing with the LBL Bevatron and the CERN ISR in the 1970s, up to the more recent LANL PSR, KEK Photon Factory, AGS Booster, KEKB CERN SPS, CERN PS, PEP-II, FNAL Main Injector, ORNL SNS, Cornell CESR-TA, and LHC, electron-cloud related beam instabilities have been observed at all storage rings operating with intense positively charged beams of protons or positrons. Figure 1 illustrates some of the historical observations.

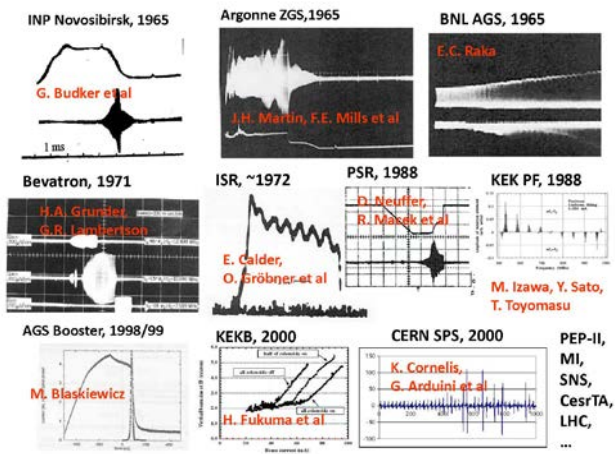

Figure 1: Historical observations .of electron-cloud effects on proton or positron storage rings [1,2,3,4,5,6,7,8,9,10,11,12,13,14,15,16].

Figure 2 presents a schematic sketch of electron-cloud build up in the beam pipe of the large Hadron Collider (LHC), based on Ref. [17]. The LHC is the first proton storage ring with significant synchrotron radiation and appreciable photon critical energy (44 eV). Photo-electrons emitted at the time of the bunch arrival are accelerated in the electrical field of the passing bunch and acquire kinetic energies up to 200 eV, so that they produce secondary electrons when they hit the wall on the other side of the beam pipe. The secondary electrons, with initially much lower energy, distribute inside the beam pipe, and are accelerated during the following bunch passage, with even higher energy gain. Both copious photoelectron production and the beam-induced multipactoring process for an average secondary electron yield larger than 1 lead to an avalanche-like build up in the beam vacuum chamber, which finally saturates due to the repelling space charge field of the "electron cloud" itself.

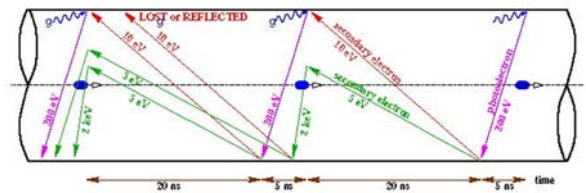

Figure 2: Schematic sketch of electron-cloud build up in an LHC beam pipe (Courtesy Francesco Ruggiero, 2000).

Figure 3 shows empirically found electron-cloud instability thresholds in terms of bunch population $N_b$ (bunch charge) as a function of the bunch spacing $s_b$ – the blue plotting symbols. Added in red are the corresponding parameters of a couple of proposed future projects: the "ultimate LHC" and the damping rings of several either historical or current linear collider projects. For a given storage ring the observed instability threshold varies approximately linearly with the bunch spacing, namely $N_{b,thr} \sim s_b$. It is worth highlighting that a simple multipactoring formula [18] predicts a rather different scaling $N_{b,thr} \sim 1/s_b$.

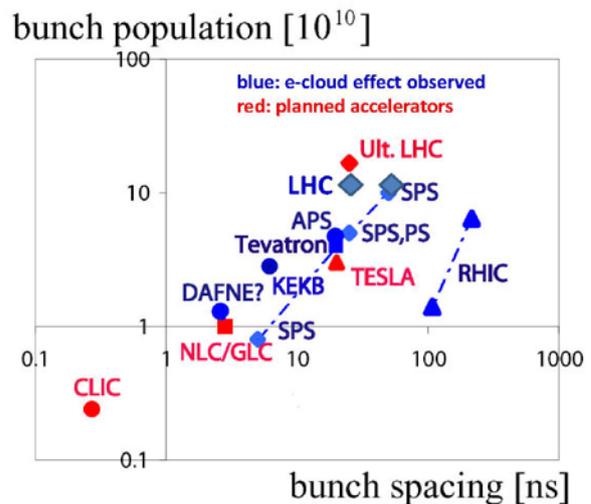

Figure 3: Bunch population at the electron-cloud instability threshold as a function of bunch spacing [19].

Electron-cloud effects observed in accelerators include:
- vacuum pressure rise;
- multi-bunch and single-bunch instabilities;

- incoherent emittance growth & beam loss;
- heat load, in particular inside cold superconducting magnets; and
- perturbation to beam diagnostics

## LHC STRATEGY AGAINST E-CLOUD

After the potential threat of an electron cloud had been realized, e.g. [17], the following mitigation strategy was developed under the leadership of Francesco Ruggiero:

- in the warm sections (20% of circumference) the beam pipe was **coated by TiZrV getter** developed at CERN, which is characterized by low secondary emission; if an electron-cloud occurs nevertheless, the residual-gas ionization by electrons (high cross section ~400 Mbarn) aids in pumping & the pressure will even improve;
- in the cold arc the outer wall of the Cu-coated LHC beam screen (at 4-20 K, installed inside the 1.9-K cold bore) is equipped with a **sawtooth surface** (30 μm steps over 500 μm period) to reduce the specular photon reflectivity to ~2% or less so that photoelectrons are almost exclusively emitted from the outer wall, where they are confined by the strong dipole field;
- the pumping slots in the beam screen are **shielded.** so as to prevent electron impact on the cold magnet bore;
- in addition one has to rely on **surface conditioning** ('scrubbing', i.e. a reduction of the secondary emission yield with continued electron bombardment) as part of the LHC commissioning strategy; as a last resort, if the scrubbing does not sufficiently reduce the secondary emission yield, doubling or tripling the bunch spacing would always suppress the e-cloud heat load to an acceptable level.

Some of these points are illustrated in the following figures. Figure 4 shows the simulated heat load per meter length in the LHC arcs for the nominal 25-ns bunch spacing as a function of bunch population (design value $N_b=1.15 \times 10^{11}$). Figure 5 presents details of the LHC beam screen in the cold arcs, including the location of the "sawtooth", cooling tubes, pumping slots, and electron-cloud shields. Figure 6 shows a cut through a prototype sawtooth surface. The conditioning of a colaminated Cu surface with 500 eV electrons as measured in the laboratory can be seen in Fig. 7, and the "scrubbing" of LHC sawtooth Cu chamber in situ with photo-electrons of critical energy 194 eV and a 100-V biasing voltage at EPA in Fig. 8. The conditioning by the electron cloud itself has been verified with Cu samples in the SPS (See Fig. 9). Also the conditioning at cryogenic temperatures was demonstrated, in the laboratory (Fig. 10).

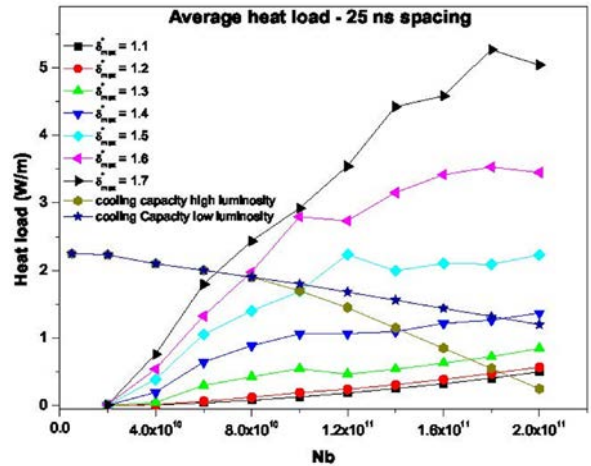

Figure 4: Simulated heat load per unit length in the LHC arcs as a function of bunch population, for different values of the maximu msecodnary emission yield $\delta_{max}^*$, together with the maximum cooling capacity available at low and high luminosity operation [20].

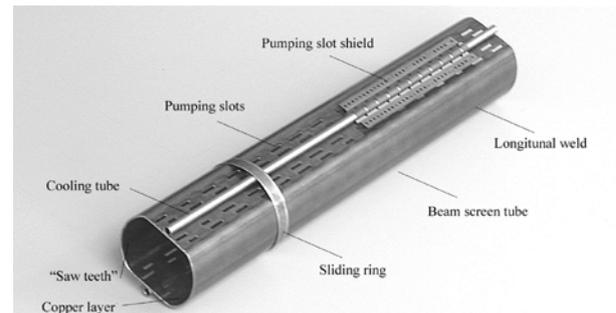

Figure 5: Conceptual design of LHC beam screen in the cold arcs [21].

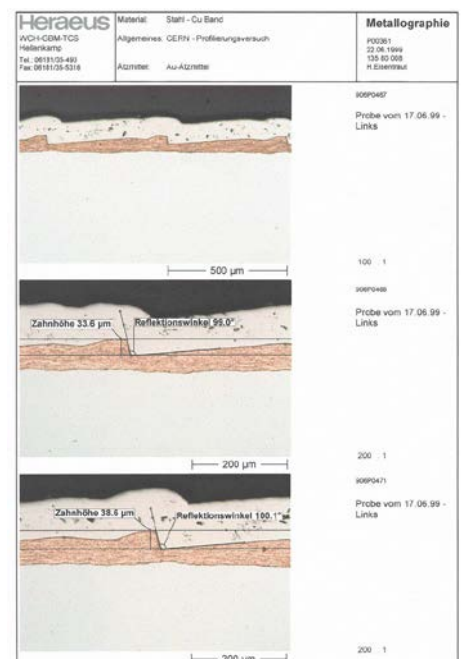

Figure 6: Sawtooth pattern imprinted on LHC vacuum chamber prototype (Courtesy Ian Collins) [22].

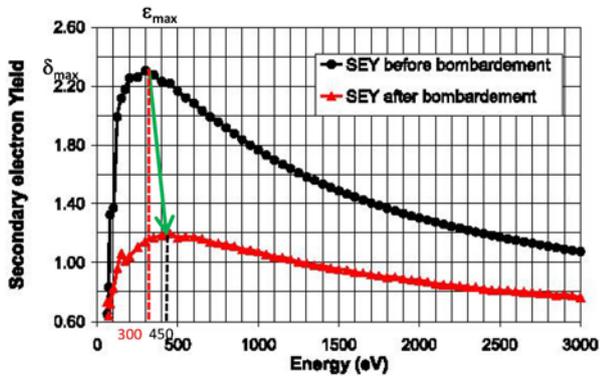

Figure 7: Variation of SEY as a function of the primary electron energy, for a sample of copper colaminated on stainless steel, before and after bombardment with 500 eV electrons, corresponding to a dose of $5 \times 10^{-3}$ C/mm$^2$ [23].

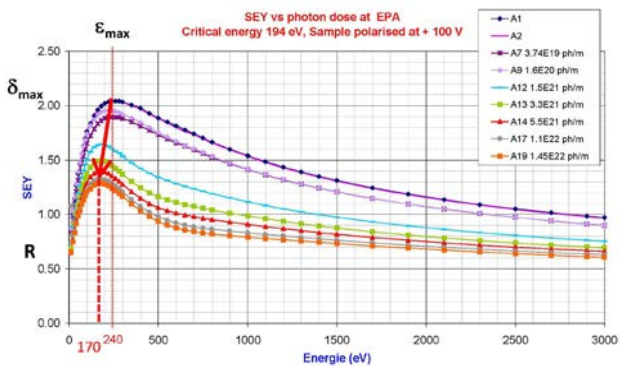

Figure 8: In-situ conditioning of the secondary emission yield for an LHC sawtooth Cu chamber with photo-electrons at EPA [24].

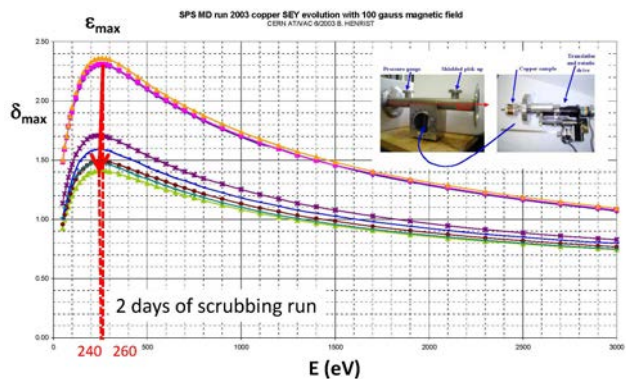

Figure 9: In-situ conditioning of the secondary emission yield for a Cu sample installed at the SPS with a beam-induced electron cloud [25].

It is interesting to note that in Fig. 7 the primary energy at which the secondary emission yield is maximum, called $\varepsilon_{max}$, increases with surface conditioning, while in Figs. 8-9 it decreases and in Fig. 10 its evolution is non-monotonic.

A further unresolved aspect pertains to the probability $R$ that an incident low-energy electron is elastic reflected when it hits the surface. In Fig. 10 the probability of elastic e- reflection $R$ for *copper* at cryogenic temperature seems to approach 1 and to be independent of $\delta^*_{max}$.

Another value for the low-energy elastic reflection probability $R$, namely $R \sim 0.2\text{-}0.5$ was obtained in a different way, by benchmarking electron cloud simulations against the electron flux seen on SPS e-cloud monitors, observed with different beam filling patterns [27]. These SPS measurements were done using an unbiased "strip detector" with copper strips located below a perforated *stainless stee*l vacuum chamber.

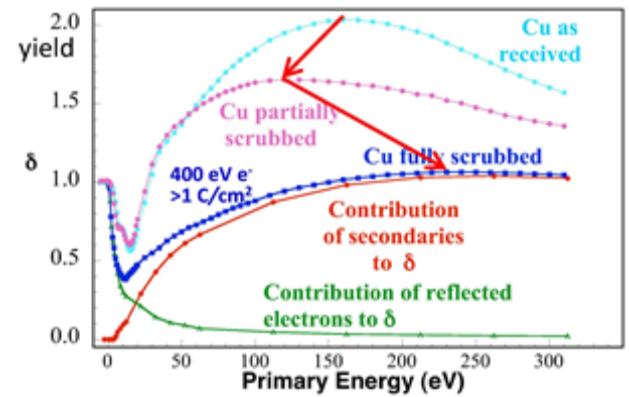

Figure 10: Conditioning of Cu sample at 10 K in special laboratory test [26].

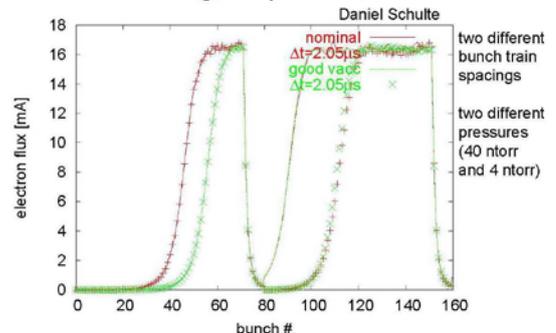

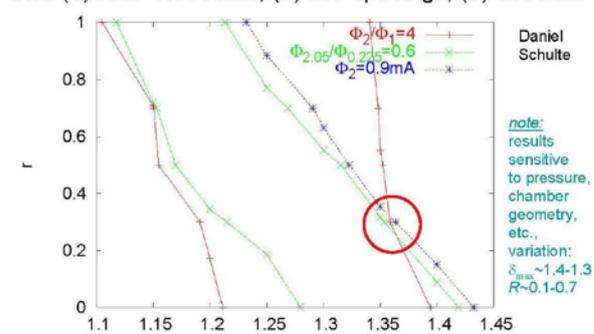

Figure 11: Benchmarking e-cloud simulations: e- flux for different spacings of two SPS bunch trains and for two different vacuum pressures [top]; measured flux ratios & flux in simulated $R\text{-}\delta_{max}$ plane [bottom] (D. Schulte) [27].

One goal of ECLOUD12 was a critical review and a summary of the experimental observations (machine and laboratory) with respect to $R$ and $\varepsilon_{max}$ values and their evolution during conditioning.

## PAST ECLOUD WORKSHOPS

The following is a list, with links, of e-cloud related workshops, which began in earnest about 15 years ago:

**Mini-Symposium on Photoelectron and Ion Instabilities at PAC'97** (Proceedings: 4 MB, PDF format), eds. J. Rogers and E. Camdzic (Cornell LNS report CLNS 97-1487, May 1997).

**MBI97**: International workshop on multibunch instabilities in future electron and positron accelerators, Tsukuba, KEK, 15 - 18 July 1997, (ed. Y.-H. Chin), KEK Proceedings 97-17

8th Advanced ICFA Beam Dynamics **Mini Workshop on Two-Stream Instabilities in Particle Accelerators and Storage Rings**, Santa Fe, February 16-18, 2000 (org. by K. Harkay and R. Macek)

International workshop on **Two-Stream Instabilities in Particle Accelerators and Storage Rings**, Tsukuba, 11-14 Sept. 2001

**ECLOUD'02**: **Mini-Workshop on the Electron-Cloud Simulations for Proton and Positron Beams**, CERN, Geneva, April 15-18, 2002. Full program and slides from the talks . Proc. CERN-2002-001

13th ICFA Beam Dynamics Mini-Workshop on **Beam Induced Pressure Rise in Rings**, Brookhaven National Laboratory, December 9-12, 2003.

**ECLOUD'04** : **31st ICFA Advanced Beam Dynamics Workshop on Electron-Cloud Effects**, Napa, California, April 19-22, 2004, Proc. CERN-2005-001

**CARE-HHH-APD workshop "HHH-2004"**, CERN, 8-11 November 2004, CERN-2005-006

**ECL2: Joint CARE-HHH, CARE-ELAN and EUROTeV Workshop on Electron Cloud Clearing**, CERN, 28 February-2 March 2007, **Proceedings CERN-AB-2007-064-ABP** (also CARE-Conf-07-007-HHH, CARE/ELAN document-2007-004 and EUROTEV-Report-2007-060)

**ECLOUD'07**: International Workshop on Electron-Cloud Effects, Daegu, Korea, 9-12 April 2007

**MulCoPim'08**: Valencia, 24-26 September 2008

**ECM'08:** CARE-HHH-APD Mini-Workshop on Electron-Cloud Mitigation, CERN, 20-21 November 2008, CARE-Conf-08-031-HHH

**AEC'09:** Topical EuCARD-AccNet workshop on anti-electron-cloud coatings that require no activation, CERN 12-13 October 2009

**ECLOUD'10** : 49th ICFA Advanced Beam Dynamics Workshop on Electron Cloud Physics, Cornell, Ithaca, 8-12 October 2010

**CERN-GSI workshop on electron cloud**, CERN, 7-8 March 2011, EuCARD-REP-2011-005

**MulCoPim'11**: International workshop on Multipactor, Corona & Passive Intermodulation, 21-23 Sept.'11

**ECLOUD'12:** INFN-LNF/INFN-Pisa/LER/EuCARD-AccNet Joint workshop, Elba, Italy, 5-9 June 2012

The workshop names, locations and years are illustrated in Fig. 12. The workshops cluster in Western Europe, the United States, and Japan/Korea region. Not a single e-cloud workshop has yet been organized in the Southern hemisphere.

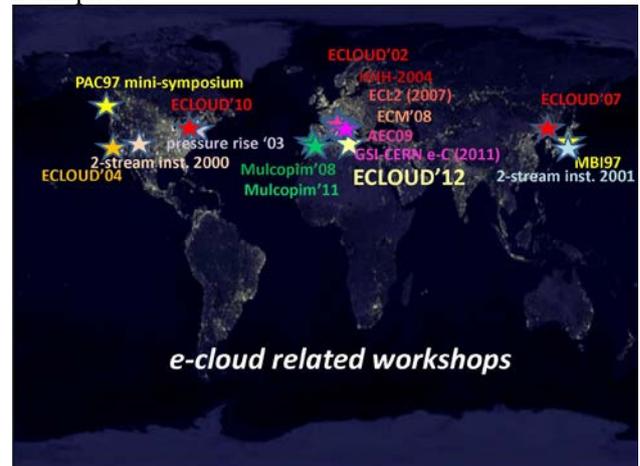

Figure 12: Locations and years of e-cloud related workshops.

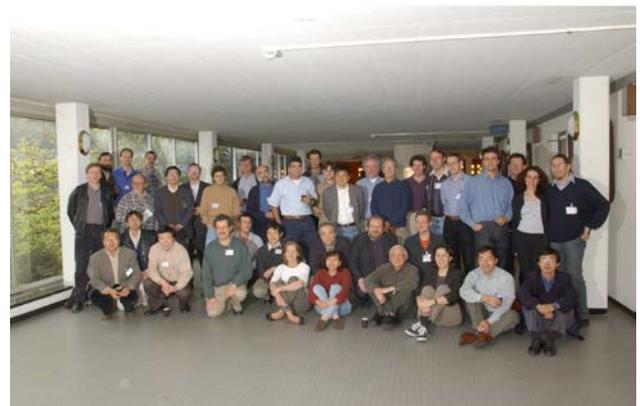

Figure 13: Group photo of ECLOUD'02 at CERN.

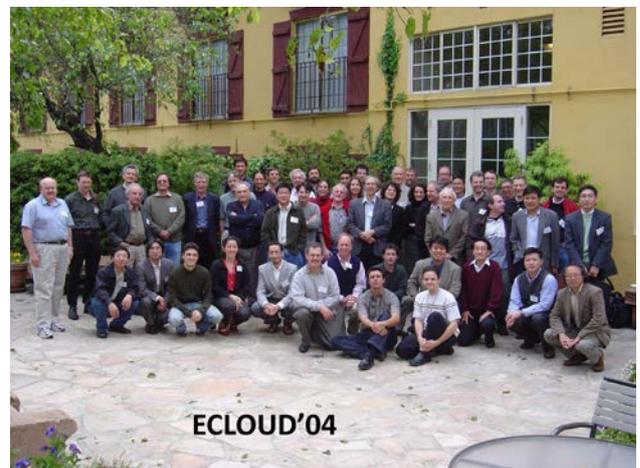

Figure 14: Group photo of ECLOUD'04 in Napa.

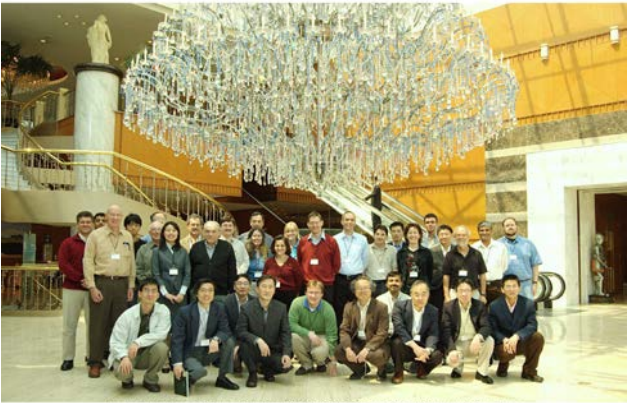

Figure 15: Group photo of ECLOUD'07 in Daegu.

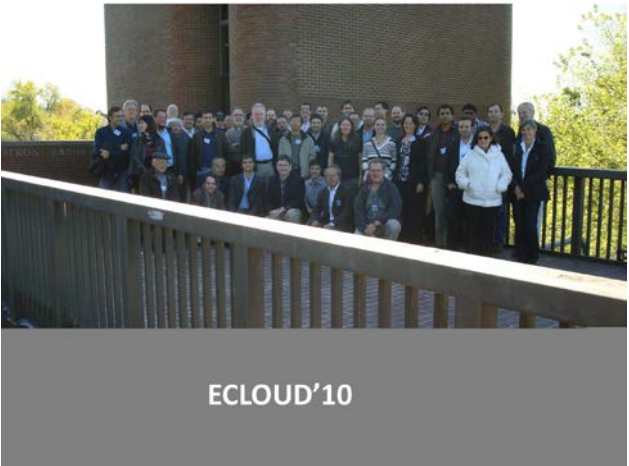

Figure 16: Group photo of ECLOUD'10 in Cornell.

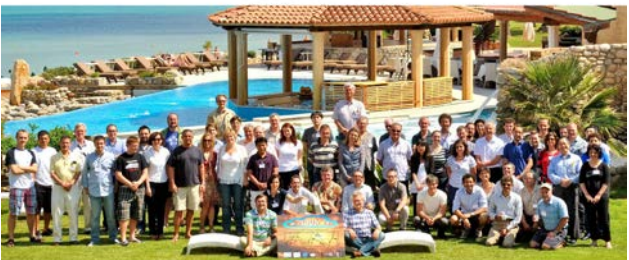

Figure 17: Group photo of ECLOUD'12 on La Biodola.

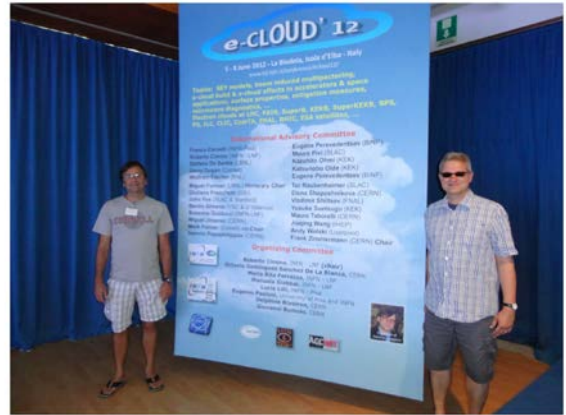

Figure 18: Photo of ECLOUD'12 chairs on La Biodola.

In particular there have been 5 ECLOUD workshops so far (2 in Europe, 2 in the US, and 1 in Asia – illustrated in Figs. 13-18), 5 topical e-cloud workshops organized in the frameworks of CARE-HHH (2004-2008) and EuCARD-AccNet (2009-2013), and two MulCoPim workshops organized jointly with, and primarily by, ESA-ESTEC and associated spacecraft organizations.

## WORKSHOP IMPACT

The first ECLOUD workshop, ECLOUD'02, revealed substantial synergies with the plasma physics & plasma acceleration communities. This has led to interesting collaborations, e.g. between CERN and USC, and to joint modeling efforts. As an example Fig. 19 shows the simulation of an LHC bunch passing through an e- plasma using the QUICKPIC code originally developed for plasma-wakefield acceleration.

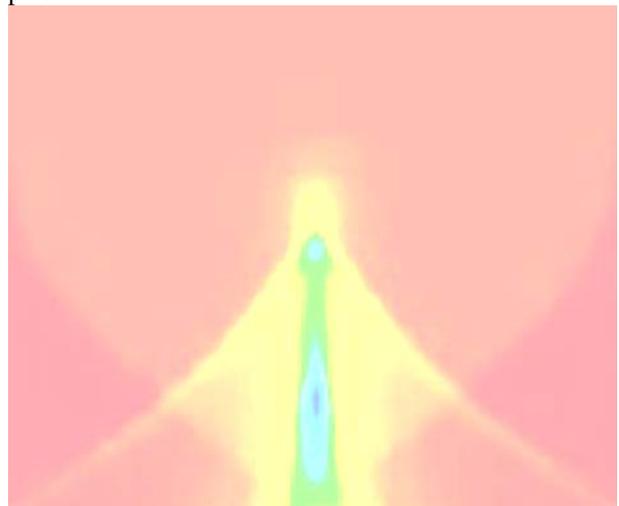

Figure 19: Simulation of LHC bunch passing through e-plasma using the QUICKPIC code (color indicates e-density) [28].

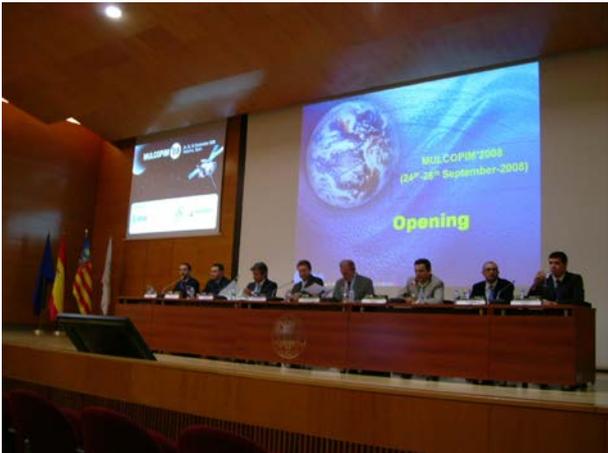

Figure 20: Opening session at MulCoPim'08 workshop in Valencia.

Synergies with the satellite community began to be realized and exploited following early contacts with Shu Lai, then at Harwood military base, and the ESA-ESTEC MulCoPim'08 workshop in Valencia. Figure 21 shows F. Caspers and the author visiting the ESTEC laboratory in The Netherlands a few months after MulCoPim'08 for discussing possible collaborations with D. Raboso.

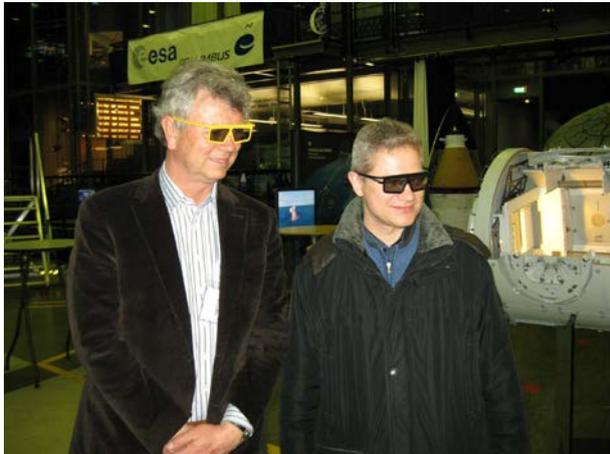

Figure 21: F. Caspers (CERN) and the author visiting ESA/ESTEC in February 2009.

The recognition of electron cloud as an important limitation for accelerator operation, in particular for the LHC, is reflected in its appearance on the internet, as illustrated in Fig. 23.

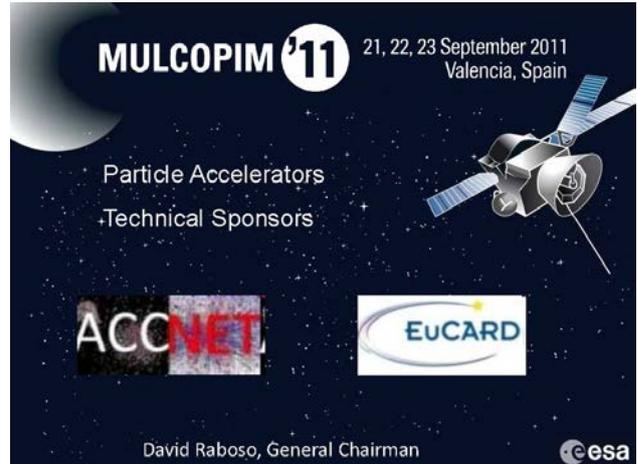

Figure 22: Opening slide at MulCoPim'11 from David Raboso, highlighting EuCARD-AccNet as a "Technical Sponsor."

Figure 23: Occurrences of "electron cloud" in google scholar.

## E-CLOUD CODE DEVELOPMENT

The first modern electron-cloud simulation code was developed in 1995 by Kazuhito Ohmi [29]; see Fig. 24. First e-cloud simulation for PEP-II were presented by Miguel Furman and Glen Lambertson in 1996 [30]; Fig. 25, and the first electron-cloud simulation for the LHC by Frank Zimmermann in early 1997 [17], Fig. 26.

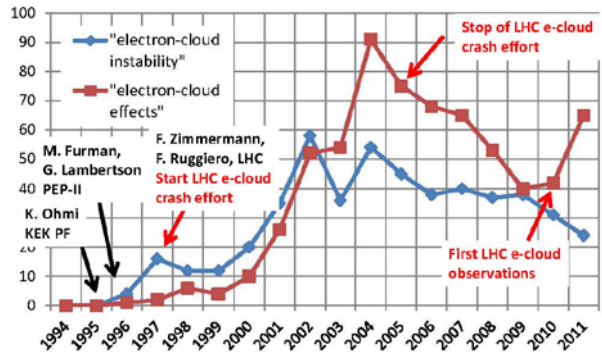

Figure 24: Reporting first electron-cloud simulations overall [29].

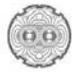

Figure 25: Reporting first electron-cloud simulation for PEP-II [30].

Figure 26: Reporting first electron-cloud simulations for the LHC [17].

Many other codes have been developed later. One distinguishes electron-cloud "build-up" codes, like POSINST, ECLOUD, CSEC, CLOUDLAND, and Factor2, instability codes, like PEHTS or HEADTAIL, incoherent codes such as MICROMAP or CMAD, and combined codes, like WARP-POSINST or VORPAL. In addition, closely related (for modeling the photoelectron emission) tools are photon tracking codes such as PHOTON or SYNRAD3D. Combined codes, such as WARP-POSINST, come closest to the goal of a "complete" electron-cloud simulation programme, defined in 2004 (Fig. 27). Namely WARP and POSINST have recently been further integrated, enabling fully self-consistent simulation of e-cloud effects: build-up & beam dynamics. Figures 29 and 30 show results from a WARP-POSINST enabled first direct simulation of a train of 3x72 bunches in the CERN SPS, accomplished by using 9,600 CPUs on a Franklin supercomputer (NERSC) [32]. Figure 30 indicates a substantial density rise in the tails of the batches between turn 0 and turn 800.

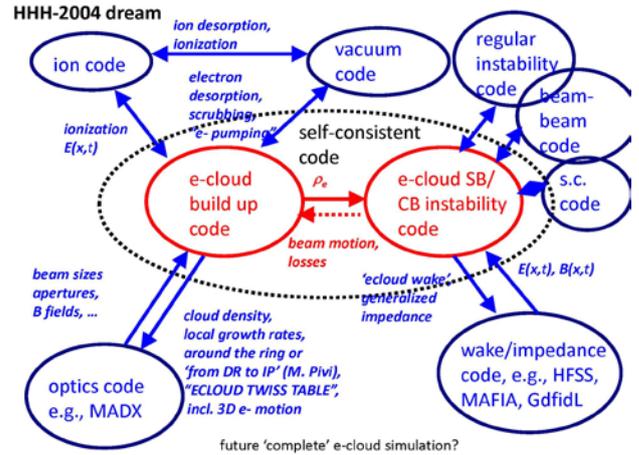

Figure: 27: A future "complete" electron-cloud simulation as conceived at CARE-HHH 2004 [31].

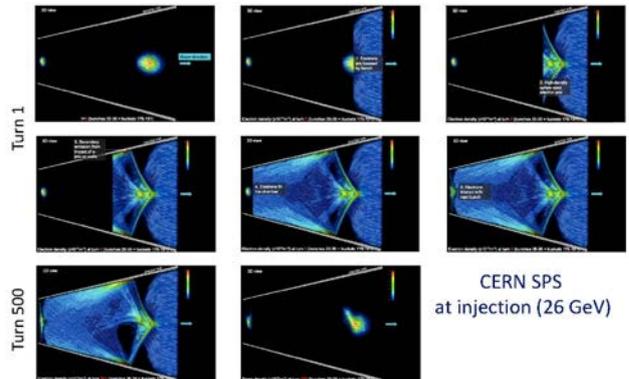

Figure 28: 3-D view snapshots of one bunch in an SPS bunch train interacting with an electron cloud generated by the preceding bunches, on turn 0 (first two rows) and on turn 500 (last row). This WARP-POSINST simulation considered 3x72 LHC-type bunches in the SPS at injection energy (26 GeV/c) [32].

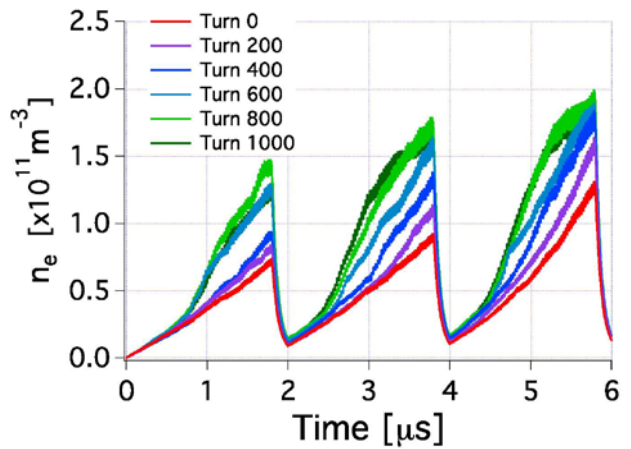

Figure 29: Average electron cloud density history at a fixed station, from the same simulation as the snapshots in Fig. 28 [32].

## FRANCESCO RUGGIERO

Francesco Ruggiero (1957-2007) was a brilliant accelerator physicist, inventive researcher, great collaborator, excellent mentor and true gentleman. In 1985 he received a PhD in accelerator physics from the prestigious Scuola Normale Superiore di Pisa. After participating in the commissioning of LEP, he made numerous invaluable contributions to the design of the LHC, in particular on collective effects, machine impedance and beam–beam interaction. Figure 30 shows him in 1996. In 1997, Francesco launched an important remedial electron-cloud crash programme for the LHC. Later he became leader of the accelerator-physics group in the former CERN SL Division. Since 2000 Francesco drove the LHC accelerator upgrade studies, for example, by coordinating the CARE-HHH network. Under his wonderful and caring guidance many bright young accelerator physicists were trained or recruited at CERN. Francesco was full of passion and energy, often working until dawn. His open mind, his love for physics, his friendliness and his humour will never be forgotten.

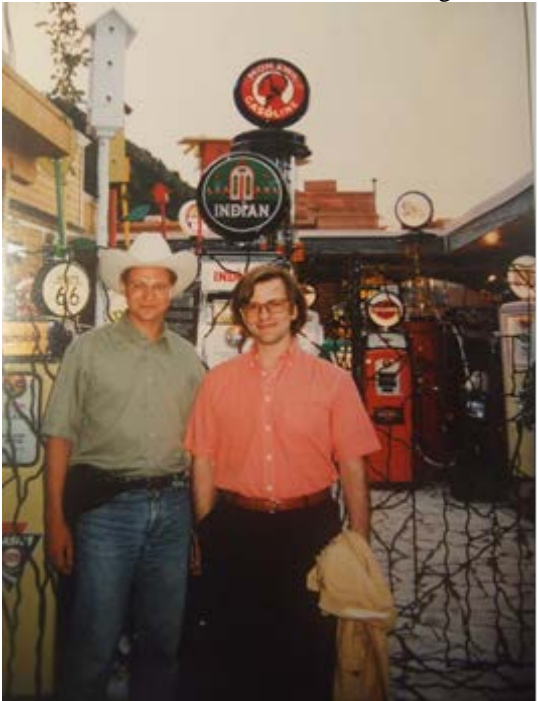

Figure 30: Francesco Ruggiero and the author in 1996, at Snowmass Colorado.

The e-cloud accomplishments either directly from, or inspired by, Francesco Ruggiero's LHC crash program were summarized by Miguel Furman for the CARE-HHH BEAM'07 workshop (which included a Francesco Ruggiero Memorial Session) [33]:

- Careful measurements of quantum efficiency & SEY in technical materials
- Identification of TiZrV as a novel low-SEY coating
- Development & deployment of several types of in-situ electron detectors
- Measurement of correlation of vacuum pressure with electron activity
- Development of new mitigation mechanisms (eg., grooved surfaces, high chromaticity mode, multibunch feedback for SPS in x-plane,…)
- First observations of the EC with LHC beam in SPS (1999) and PS (2001)
- Practical demonstration of self-conditioning of the ECE at SPS (~few days)
- Measurement of EC flux and energy spectrum at SPS and RHIC with these detectors
- Development of careful secondary emission models
- Understanding via analytical models
- Great developments in simulation codes, validation, and benchmarking
- Prediction of ECE density and power deposition for LHC
- Investigation of ECEs in other types of machines (eg., heavy-ion linacs)
- Investigation of severity of ECE against fill pattern, bunch intensity, etc

## ECLOUD12 TOPICS

Topics to be addressed at ECLOUD12 included SEY models, e-cloud build & e-cloud effects in accelerators & space applications, beam induced multipactoring, surface properties, mitigation measures, microwave diagnostics, as well as, more specifically, electron clouds at LHC, FAIR, SuperB, KEKB, SuperKEKB, SPS, PS, ILC, CLIC, Cesr-TA, FNAL, RHIC, and ESA satellites.

## ACKNOWLEDGEMENTS

We acknowledge the support of the European Community-Research Infrastructure Activity under the FP7 Research Infrastructures project EuCARD, grant agreement no. 227579.